\documentclass[superscriptaddress,aps,pra,twocolumn,showpacs,nofootinbib,longbibliography]{revtex4-1}
\usepackage{amsmath,amssymb,amsthm}
\usepackage{easybmat}
\usepackage[colorlinks=true,citecolor=blue,urlcolor=blue]{hyperref}
\usepackage[pdftex]{graphicx}
\usepackage{times,txfonts}
\usepackage{braket}
\usepackage{color}
\usepackage{natbib}
\newcommand{\be}{\begin{equation}}
\newcommand{\ee}{\end{equation}}
\newcommand{\ba}{\begin{eqnarray}}
\newcommand{\ea}{\end{eqnarray}}
\newcommand{\ketbra}[2]{|#1\rangle \langle #2|}

\begin{document}
	
	\title{Quantum correlations and speed limit of central spin systems}

	\author{Devvrat Tiwari\textsuperscript{}}
	\email{devvrat.1@iitj.ac.in}
	\author{K.G Paulson\textsuperscript{}}
	\email{paulsonkgeorg@gmail.com }
	\author{Subhashish Banerjee\textsuperscript{}}
	\email{subhashish@iitj.ac.in }
			\affiliation{Indian Institute of Technology Jodhpur-342030, India\textsuperscript{}}



\date{\today}

\begin{abstract}
	In this article, we consider single, and two-qubit central spin systems interacting with spin baths and discuss their dynamical properties. We consider the cases of interacting and non-interacting spin baths and investigate the quantum speed limit (QSL) time of evolution. The impact of the size of the spin bath on the quantum speed limit for a single qubit central spin model is analyzed. We estimate the quantum correlations for (non-)interacting two central spin qubits and compare their dynamical behaviour with that of QSL time under various conditions. We show how QSL time could be availed to analyze the dynamics of quantum correlations.
\end{abstract}

\maketitle

\section{\label{sec:Intro} Introduction }  
A quantum system's evolution almost inevitably involves its ambient environment's influence. This is handled in a consistent manner by the theory of open quantum systems \cite{ bruer-petrrucione,sbbook}. Open system ideas find use in a wide range of fields ranging from quantum optics \cite{Louisell,GSAgarwal}, to condensed matter \cite{weiss,caldeira-leggett1983}, quantum statistical physics \cite{GrabertSchrammIngold,sbsterngerlach,sbqbm}, quantum chemistry \cite{reactioncoordinaterefs1, reactioncoordinaterefs2, reactioncoordinaterefs3}, and quantum biology \cite{plenio}. In the last decade, its applications to the field of quantum information have seen phenomenal growth \cite{sbindranil,sbsrik,sbgeorge,sbgp,sbchandruparrando,sbsamyapati,pillobassano}. The quantum system's coupling to the reservoir (bath or environment)  plays a significant role in deciding the nature of its dynamics, Markovian or non-Markovian \cite{blp,rhp,hcla,sss,lidar}, and the usefulness of quantum resources for technological advancements. 

The dynamics of quantum correlations under the (non-)Markovian environment have been extensively studied \cite{sbjavidsupriyo, franco_quant_corr_nm, paulson2021hierarchy, Atta_Rehman_quant_corr}. The non-Markovian behaviour, arising from, say, a strong system-bath coupling, can delay the decay and sometimes source the revival of quantum correlations \cite{pradeep, wang_system_bath}. The evolution of the system of interest can change due to the nature of the bath, and this could be witnessed from the dynamics of quantum speed limit (QSL) time~\cite{deffnerreview, Pfeifer, Frey2016}. Energy-time uncertainty reveals the bound on the speed of the evolution of quantum states~\cite{mandelstam1945}. A bound on the speed of dynamical evolution was initially derived for the dynamics between the orthogonal states for isolated systems~\cite{MARGOLUS}. QSL time was later extended to take into account the case in which the system does not evolve to an orthogonal state~\cite{giovannetti_qsl}. Furthermore, the speed limit for driven quantum systems that are valid for arbitrary initial and final states, as well as for arbitrary unitary driving, was derived~\cite{Deffner_2013}. QSL time for the evolution between arbitrary states in open quantum systems finds many potential applications~\cite{Aifer_2022, paulson2021effect, paulson2022quantum, Minglun_2017} in the field of quantum information and computation and is an active research topic~\cite{Campaioli_2022, non-uniform, qsl_appl_2}.

In general, comprehending the subtlety of the environment's impact on the quantum system is a non-trivial task. In this work, we discuss the central spin system~\cite{Breuer_spin_star, wen-bin}, both single as well as double spin systems~\cite{mukhopadhyay2017, two-qubit-central-spin}, interacting with spin bath(s) and investigate the corresponding evolution under various conditions. The class of spin bath models \citep{Prokof_ev} are of immense interest in the quantum theory of magnetism \citep{parkinson}, quantum spin glasses \citep{Rosenbaum}, quantum batteries~\cite{Liu_quantum_battery}, NV-center~\cite{Hanson_spin_bath}, theory of conductors and superconductors \citep{leggett-two-state}. In a previous work~\cite{two-qubit-central-spin} pertaining to two central spins immersed in spin baths with interacting bath spins, the dynamical map of the reduced system using Holstein-Primakoff transformation was derived, and its non-Markovian behaviour studied.
Here we investigate the QSL time and the dynamics of quantum correlations of the central spin system interacting with the spin bath. It is worth mentioning that if we use the Holstein-Primakoff transformation, the Hamiltonian describing the collective behaviour of $N$ interacting spins can be mapped to a bosonic one. With homogenous interactions and in the limit of a large number of bath spins, the central spin model transforms into a nonlinear Jaynes-Cummings model. This opens the opportunity to study
more complex quantum devices like a quantum thermal diode~\cite{diode}. Here we have exactly solved the single as well as two central spin systems numerically, taking into account the impact of bath spin interactions.
This is used to characterize the quantum statistical mechanics of the central spin model. The presence of bath spin interactions is seen to contribute significantly to the dynamics. This sets apart the spin bath from its bosonic counterpart, where the bath is composed of independent modes of harmonic oscillators. For the two central spin systems, we will see that it is fruitful to consider the dynamics as local or global depending on the absence or presence of the spin-spin interaction. Quantum correlations, in particular, entanglement and discord, are calculated, and their connection with QSL time under different conditions is discussed. 

The present work is structured as follows: In Sec. \ref{sec:Model}, we discuss the details of the single and two-qubit central spin model under the influence of local and global baths. Section \ref{sec:Characterization} studies the characterization of the central spin model and the impact of bath spin interactions. Investigation of QSL time for the single qubit central spin system is given in Sec. \ref{sec:QSL-time}. The dynamics of quantum correlations and their connection with QSL time for the two-qubit central spin system is discussed in Sec. \ref{sec:quantum-correlations}, followed by the conclusion.

\begin{figure}[h]
	\includegraphics[width=1\columnwidth]{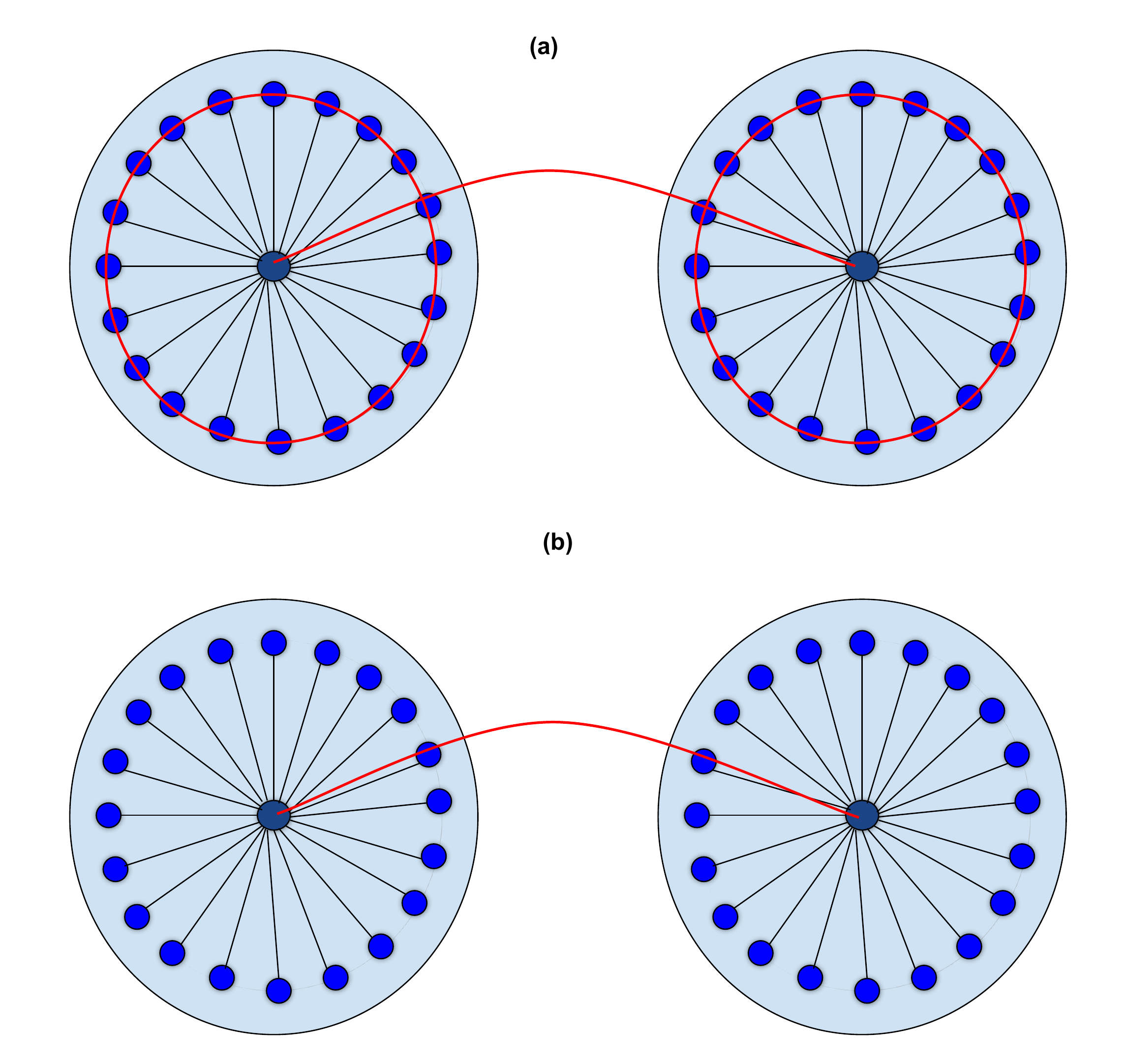}
	\caption{Schematic diagram for two central spins immersed in spin baths. In (a), we have interacting bath spins and in (b), we have non-interacting bath spins. Central spins interact with respective spin baths uniformly.}
	\label{fig-schematic_diagram}
\end{figure}

\section {\label{sec:Model} The Model}
In this section, we present the central spin model, that is, a spin-1/2 particle (central spin) coupled to a thermal bath of spin-1/2 particles. The central spin interacts uniformly with the bath spins. We discuss the dynamics of single and two-qubit central spin models. The bath spins in these models can be interacting as well as non-interacting. A schematic diagram depicting various conditions is shown in Fig.~\ref{fig-schematic_diagram}. Figure~\ref{fig-schematic_diagram}(a) represents the case of interacting bath spins while (b) corresponds to non-interacting bath spins. If the red line connecting the two central spins is removed, they represent the analogous single central spin models.

\subsection{Single qubit central spin model}
Here we consider a single central spin-1/2 particle interacting uniformly with a bath of spin-1/2 particles. We divide the single qubit, central spin model, into two categories, one in which the bath spins interact and the other with no interaction among bath spins.

\subsubsection{Single qubit central spin model with interacting bath spins}
Below we take into account a single central spin interacting uniformly and centrally with a spin bath, and the bath spins also interact with each other. The Hamiltonian of this system is given by
\begin{equation}
	\begin{aligned}
		H_{\text{single}} &= H_S + H_E + H_{SE}, \\
		&= \frac{\hbar}{2}\omega_0\sigma_0^z + \frac{\hbar\omega}{2N}\sum_{i=1}^N\bigg\{\frac{1}{2}\sum_{\substack{j=1 \\ j\ne i}}^N(\sigma^x_i\sigma^x_j + \sigma^y_i\sigma^y_j) + \sigma^z_i\bigg\} \\
		&+ \frac{\hbar\epsilon}{2\sqrt{N}}\sum_{i=1}^N(\sigma_0^x\sigma_i^x + \sigma_0^y\sigma_i^y),
		\label{total-H-single spin}
	\end{aligned}  
\end{equation}
where $H_S$, $H_E$, and $H_{SE}$ are the system, environment and interaction Hamiltonians, respectively. Here $\sigma_0^k$ $(k = x, y, z)$ represents the Pauli spin matrices for the central spin system, and $\sigma_i^k$ are the Pauli spin matrices for the $i^{th}$ spin of the bath. $N$ represents the number of spins in the bath. The frequency of the bath and the bath-spin interaction strength are rescaled as $\omega/N$ and $\epsilon/\sqrt{N}$, respectively. The above single qubit central spin model is an extension of the model given in \cite{mukhopadhyay2017} with the added condition that the bath spins also interact with each other. Using the collective angular momentum operators, $J_k = \frac{1}{2}\sum_i\sigma^k_i$ (with $k = x, y, z, +, -$), for the bath spins, we can rewrite the interaction and bath Hamiltonians as
\begin{align}
    H_{SE} &= \frac{\hbar \epsilon}{\sqrt{N}}(\sigma^x_0J_x + \sigma^y_0 J_y), \nonumber \\
    H_E &= \hbar \omega\bigg(\frac{J_+J_-}{N} - \frac{{\mathbb I}}{2}\bigg).
    \label{single-qubit-ang-mom}
\end{align}
We assume that the initial joint system bath state is a product state, $\rho_{SE}(0) = \rho_S(0)\otimes\rho_E(0)$. We consider the initial bath state as a thermal state $\rho_E(0) = e^{-\frac{H_E}{KT}}/Z$. Here $K$ is the Boltzmann constant, $T$ is the temperature and $Z = {\rm Tr}\left(e^{-\frac{H_E}{KT}}\right)$ is the partition function.
After a unitary evolution of the joint system-bath state, the reduced state of the system is given as
\begin{equation}
    \rho_S(t) = {\rm Tr}_E\left[e^{-iHt/\hbar}\rho_{SE}(0)e^{iHt/\hbar}\right]. 
    \label{single-qubit-rho-case1}
\end{equation}
We have numerically computed the above density matrix and used the result in the subsequent sections.

\subsubsection{Single qubit central spin model with non-interacting bath spins}
We can also take into account the effect of a non-interacting bath on a single qubit central spin system. The Hamiltonian for this system is given by
\begin{equation}
	\begin{aligned}
		H'_{\text{single}} &= H_S + H'_E + H_{SE}, \\
		&= \frac{\hbar}{2}\omega_0\sigma_0^z + \frac{\hbar\omega}{2N}\sum_{i=1}^N \sigma^z_i+ \frac{\hbar\epsilon}{2\sqrt{N}}\sum_{i=1}^N(\sigma_0^x\sigma_i^x + \sigma_0^y\sigma_i^y),
		\label{total-Hp-single spin}
	\end{aligned}  
\end{equation}
The bath Hamiltonian can be rewritten using the collective angular momentum operators, $J_k = \frac{1}{2}\sum_i\sigma^k_i$ (with $k = x, y, z, +, -$), for the bath spins as 
\begin{align}
    H'_E &=  \frac{\hbar\omega}{N}J_z.
    \label{single-qubit-ang-mom2}
\end{align}
The reduced state of the system after a unitary evolution of the joint system-bath state, in this case, is
\begin{equation}
    \rho'_{S} = {\rm Tr}_E\left(e^{-iH't/\hbar}\rho'_{SE}(0)e^{iH't/\hbar}\right).
    \label{single-qubit-Kraus-case2}
\end{equation}
Here $\rho'_{SE}(0) = \rho_S(0) \otimes \rho'_E(0)$ such that $\rho'_E(0) = e^{-\frac{H'_E}{KT}}/Z$. 
\begin{figure}[h]
	\includegraphics[height=80mm,width=1\columnwidth]{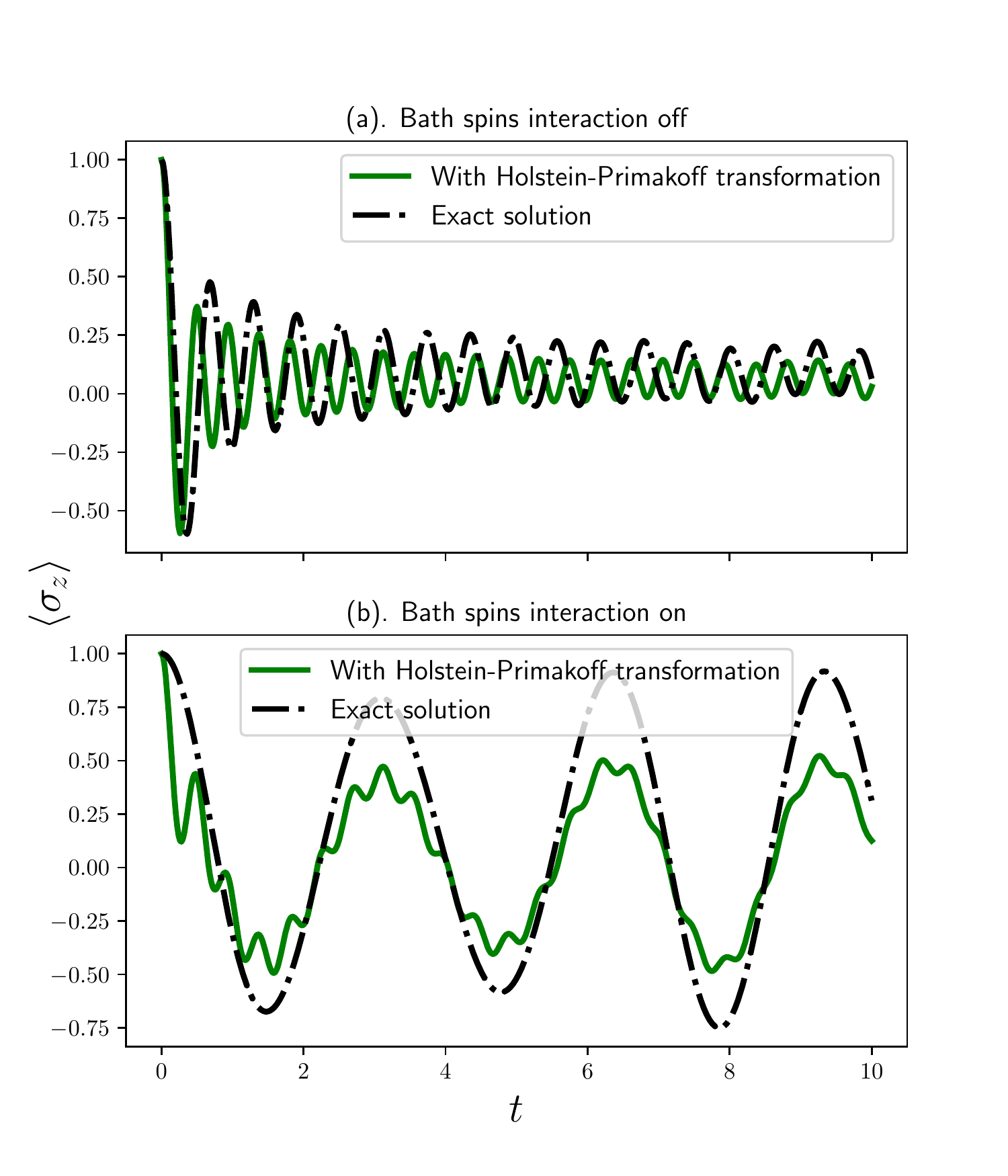}
	\caption{Variation of spin excitation $\langle\sigma_z\rangle$ for single qubit central spin models with interacting and non-interacting bath spins given in Eqs. (\ref{single-qubit-rho-case1}) and (\ref{single-qubit-Kraus-case2}), respectively, with time. The solid green lines in subplots (a) and (b) are the solutions obtained using the Holstein-Primakoff transformation. The dot-dashed black lines are the exact solutions obtained numerically. The parameters are: $\epsilon = 1$, $\omega = 2$, $\omega_0 = 2$, $N=100$, $T=1$.}
	\label{fig-diff_with_wout_hp_1_qubit}
\end{figure}

Figure~\ref{fig-diff_with_wout_hp_1_qubit} contrasts the single qubit dynamics studied from the perspective of Holstein-Primakoff transformation~\cite{mukhopadhyay2017, two-qubit-central-spin} and numerically, as in this work. Even though the general profile of both the solutions is qualitatively similar, we observe that for the interacting bath spins scenario, Fig. ~\ref{fig-diff_with_wout_hp_1_qubit} (b), the fluctuations seen in the solution obtained via the Holstein-Primakoff transformation are smoothed out in the exact solution.  This provides a motivation for going beyond the Holstein-Primakoff approximation using exact numerical methods.
\subsection{Two qubits central spin model}
We now consider two central spin-1/2 particles interacting uniformly with two separate spin baths. We divide the model into two categories, \textit{viz.} (local-)global bath scenarios depending upon the (non-)interaction between the two central spins \citep{DECORDI, two-qubit-central-spin}. Further, we also differentiate the above models based on the interaction of bath spins with each other.  
\subsubsection{Local bath scenario with interacting bath spins}
Here we consider two central spins ($A$ and $B$) with frequencies $\omega_{0\;A}$ and $\omega_{0\;B}$ interacting uniformly with their respective local baths ($N_A$ and $N_B$) with frequencies $\omega_A$ and $\omega_B$. $\epsilon_A$ and $\epsilon_B$ are the interaction strengths between each central spin and its local bath. 
We first consider the case where the spins of each bath interact among themselves, and there is no interaction between the central spins. We derive the dynamical map in this case by taking a product of the local dynamical maps for each central spin, which are obtained by tracing out the bath after a unitary evolution of the total system and bath.
If $\Lambda_A$ and $\Lambda_B$ are the dynamical maps of the evolution of the single qubit central spin systems with interacting bath spins (dynamics defined by Eq. (\ref{total-H-single spin})), the dynamics of the reduced state of the two-qubit central spin system in local bath scenario is then given by
\begin{align}
	\rho_{AB}(t) = \Lambda_A \otimes \Lambda_B \left(\rho_{AB}(0)\right). 
	\label{reduced-state-local-bath}
\end{align}
\subsubsection{Local bath scenario with non-interacting bath spins}
We next consider the case where the spins of the individual baths are non-interacting, with no interaction between the central spins. To this end, the dynamics of the reduced state are given as
\begin{equation}
    \rho_{AB}'(t) = \Lambda'_A \otimes \Lambda'_B \left(\rho'_{AB}(0)\right),
    \label{reduced-state-local-case2}
\end{equation}
where $\Lambda'_A$ and $\Lambda'_B$ are the dynamical maps of the evolution of the single qubit central spin systems with non-interacting bath spins (dynamics defined by Eq. (\ref{total-Hp-single spin})). 

The dynamical maps for each central spin were obtained in Ref.~\cite{mukhopadhyay2017} using Holstein-Primakoff transformation. However, in this work, instead of the Holstein-Primakoff transformation, we use direct numerical integration to solve Eqs. (\ref{rho-global-case1}) and (\ref{rho-global-case2}), given below.
\subsubsection{Global bath scenario with interacting bath spins}
Here each central spin is uniformly coupled to its respective bath and also interacts with each other. First, we consider the case where the spins of the individual baths interact with each other. The Hamiltonian for this particular case is given by

\begin{align}
	{\mathcal H} &= H_{S_1} + H_{S_2} + H_{S_1S_2} + H_{E_1} + H_{E_2} + H_{S_1E_1} + H_{S_2E_2}, \nonumber \\
	&=\frac{\hbar \omega_1}{2}\sigma_{01}^z + \frac{\hbar \omega_2}{2}\sigma_{02}^z + \frac{\hbar \delta}{2}(\sigma_{01}^z\otimes \sigma_{02}^z) \nonumber \\ 
	&+ \frac{\hbar \omega_a}{2M}\sum_{i=1}^M \big[\frac{1}{2}\sum_{\substack{j=1 \\ j\ne i}}^M (\sigma_{i1}^x\sigma_{j1}^x + \sigma_{i1}^y\sigma_{j1}^y) + \sigma_{i1}^z\big] \nonumber \\
	&+ \frac{\hbar \omega_b}{2N}\sum_{i=1}^N\big[\frac{1}{2}\sum_{\substack{j=1 \\ j\ne i}} (\sigma_{i2}^x\sigma_{j2}^x + \sigma_{i2}^y\sigma_{j2}^y)+ \sigma_{i2}^z\big] \nonumber \\
	&+ \frac{\hbar \epsilon_1}{2\sqrt{M}}\sum_{i=1}^M(\sigma_{01}^x\sigma_{i1}^x + \sigma_{01}^y\sigma_{i1}^y) \nonumber \\
	&+ \frac{\hbar \epsilon_2}{2\sqrt{N}}\sum_{i=1}^N(\sigma_{02}^x\sigma_{i2}^x + \sigma_{02}^y\sigma_{i2}^y),
	\label{global-hamiltonian-case1}
\end{align}
where $\sigma_{il}^k$ $(k=x,y,z;\;l=1,2)$ are Pauli matrices corresponding to $i^{th}$ spin of the $l^{th}$ bath, and $\sigma_{0l}^k$ are Pauli matrices for $l^{th}$ central spin. $\omega_{a,b}$ are bath frequencies of two spin baths and $\epsilon_{1,2}$ are interaction parameters of the respective spin-bath interaction. $M$ and $N$ are the respective numbers of spins in two baths. 
We use the collective angular momentum operators, $J_{kl} = \frac{1}{2}\sum_i\sigma^k_{il}$ (with $k = x, y, z, +, -; l = 1, 2$), for the bath spins, to rewrite the interaction Hamiltonians as 
\begin{align}
    H_{S_1E_1} &= \frac{\hbar \epsilon_1}{\sqrt{M}} \left(\sigma_{01}^x J_{x1} + \sigma_{01}^y J_{y1}\right), \nonumber \\ 
    H_{S_2E_2} &= \frac{\hbar \epsilon_2}{\sqrt{N}} \left(\sigma_{02}^x J_{x2} + \sigma_{02}^y J_{y2}\right), 
\end{align}
and the bath Hamiltonians as 
\begin{align}
    H_{E_1} &= \hbar \omega_a\bigg(\frac{J_{+1}J_{-1}}{M} - \frac{{\mathbb I}}{2}\bigg), \nonumber \\
    H_{E_2} &= \hbar \omega_b\bigg(\frac{J_{+2}J_{-2}}{N} - \frac{{\mathbb I}}{2}\bigg).
\end{align}
For the initial state $\rho_{S_1S_2E_1E_2}(0) = \rho_{S_1S_2}(0) \otimes \rho_{E_1}\otimes \rho_{E_2}$, 
the reduced state of the two central spins, after a unitary evolution of the joint system-bath state, boils down to  
\begin{equation}
\rho_{S_1S_2}(t) = {\rm Tr}_{E_1E_2}\left(e^{-i{\mathcal H}t/\hbar}\rho_{S_1S_2E_1E_2}(0)e^{i{\mathcal H}t/\hbar}\right).
\label{rho-global-case1}
\end{equation}
\subsubsection{Global bath scenario with non-interacting bath spins}
Finally, we consider a similar model but with non-interacting bath spins. The Hamiltonian, in this case, is given by
\begin{align}
	{\mathcal H'} &= H_{S_1} + H_{S_2} + H_{S_1S_2} + H'_{E_1} + H'_{E_2} + H_{S_1E_1} + H_{S_2E_2}, \nonumber \\
	&=\frac{\hbar \omega_1}{2}\sigma_{01}^z + \frac{\hbar \omega_2}{2}\sigma_{02}^z + \frac{\hbar \delta}{2}(\sigma_{01}^z\otimes \sigma_{02}^z) + \frac{\hbar\omega_a}{2M}\sum_{i=1}^M \sigma^z_{i1} \nonumber \\
	&+ \frac{\hbar\omega_b}{2N}\sum_{i=1}^N \sigma^z_{i2}+ \frac{\hbar \epsilon_1}{2\sqrt{M}}\sum_{i=1}^M(\sigma_{01}^x\sigma_{i1}^x + \sigma_{01}^y\sigma_{i1}^y) \nonumber \\
	&+ \frac{\hbar \epsilon_2}{2\sqrt{N}}\sum_{i=1}^N(\sigma_{02}^x\sigma_{i2}^x + \sigma_{02}^y\sigma_{i2}^y).
	\label{global-hamiltonian-case2}
\end{align}
Here the bath Hamiltonians, $H'_{E_1}$ and $H'_{E_2}$ are given in terms of collective angular momentum operators as 
\begin{align}
    H'_{E_1} = \frac{\hbar\omega_a}{2M}J_{z1}, && H'_{E_2} = \frac{\hbar\omega_b}{2N}J_{z2}.
\end{align}
Taking a similar initial state, the reduced state of the two central spins, in this case, after a unitary evolution of the joint system-bath state, is given by
\begin{equation}
    \rho'_{S_1S_2}(t) = {\rm Tr}_{E_1E_2}\left(e^{-i{\mathcal H'}t/\hbar}\rho'_{S_1S_2E_1E_2}(0)e^{i{\mathcal H'}t/\hbar}\right).
\label{rho-global-case2}
\end{equation}
\section{\label{sec:Characterization}Characterization of Central Spin Model: Impact of Bath spin Interactions}
Here, we study the quantum statistical nature of the central spin models. We use trace distance and von Neumann entropy to identify the changes in the single qubit and the two-qubit central spin models caused by bath spin interactions.
\subsection{Trace distance}
\begin{figure}[h]
	\includegraphics[height=65mm,width=1\columnwidth]{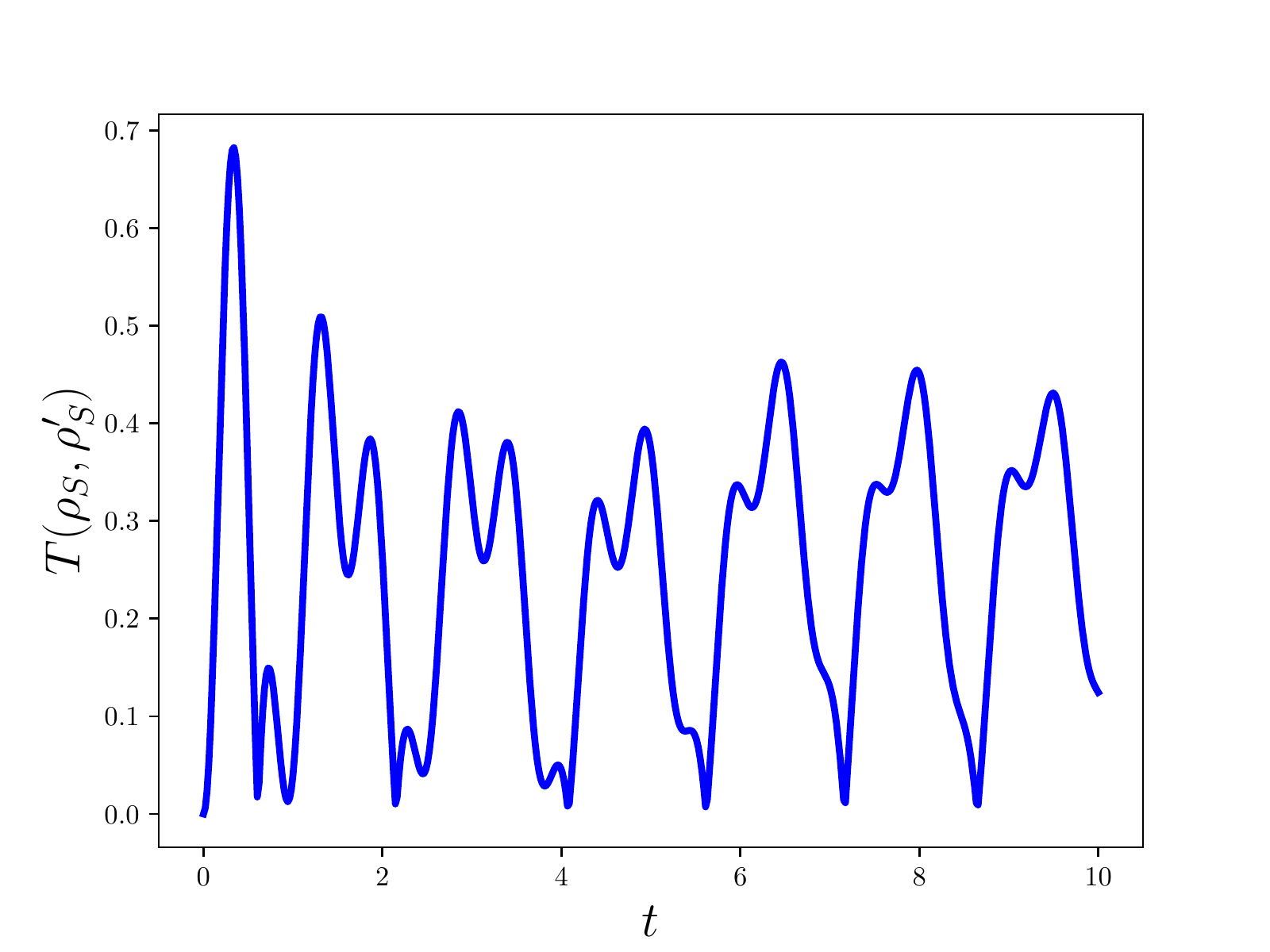}
	\caption{Variation of trace distance ($T(\rho_S, \rho'_S)$) between single qubit central spin states when interaction in bath spins is turned on $(\rho_S)$ or off $(\rho'_S)$. The parameters have following values: $\epsilon = 1$, $\omega = 2$, $\omega_0 = 2$, $N=100$, $T=1$.}
	\label{fig-trace-distance-single}
\end{figure}

We observe the effect of the interaction of bath spins over the dynamics of the central spin, both in single-qubit and two-qubit central spin models. 
Trace distance gives a measure of the distinguishability between two quantum states and is given by
\begin{equation}
    T(\rho_1, \rho_2) = \frac{1}{2}\mathrm{Tr}\sqrt{(\rho_1-\rho_2)^{\dagger}(\rho_1 - \rho_2)}.
    \label{trace-distance-def}
\end{equation}
We first calculate the trace distance between the single qubit central spin systems for the (non-)interacting bath spins. We take the initial state in both cases to be $\ket{1}$ and plot the trace distance between $\rho_S$ and $\rho_S'$ from Eq. (\ref{single-qubit-rho-case1}) and Eq. (\ref{single-qubit-Kraus-case2}), respectively, in Fig. \ref{fig-trace-distance-single}. 
The impact of bath spin interactions on the single qubit central spin system is clearly brought out and is seen to be non-monotonic.  

\begin{figure}[h]
	\includegraphics[height=65mm,width=1\columnwidth]{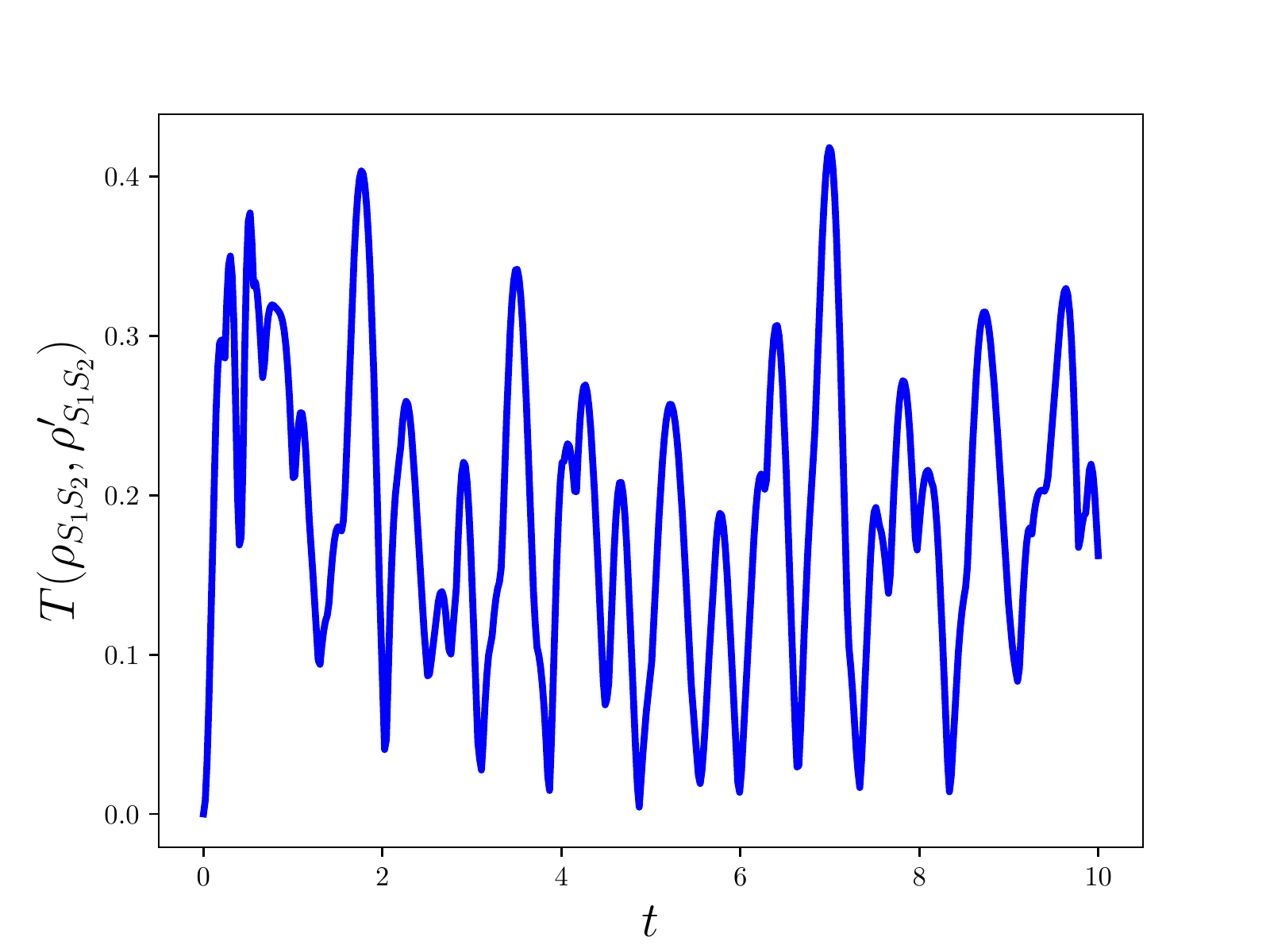}
	\caption{Variation of trace distance ($T(\rho_{S_1S_2}, \rho'_{S_1S_2})$) between two-qubit central spin states, in global bath scenario, when interaction in bath spins is turned on $(\rho_{S_1S_2})$ or off $(\rho'_{S_1S_2})$. The parameters are: $\epsilon_1 = 2.4$, $\epsilon_2 = 2.5$, $\omega_1 = 3.0$, $\omega_2 = 3.1$, $\omega_a = 2.0$, $\omega_b = 2.1$, $\delta = 4$, $M=15$  $N=15$, $T=1$.}
	\label{fig-tr-dist-2-qubit-g}
\end{figure}

For the two-qubit central spin system, we consider the initial state as $\ket{11}$ for the two-qubit central spin system. In the global bath scenario, we calculate the trace distance between the states $\rho_{S_1S_2}$ and $\rho'_{S_1S_2}$ given in Eq. (\ref{rho-global-case1}) and Eq. (\ref{rho-global-case2}), respectively. Fig. \ref{fig-tr-dist-2-qubit-g} depicts the difference in evolution between the cases, where the individual bath spins are interacting or not.  
In the local bath scenario, we calculate the trace distance between states $\rho_{AB}$ and $\rho'_{AB}$ from Eq. (\ref{reduced-state-local-bath}) and Eq. (\ref{reduced-state-local-case2}), respectively. 
\begin{figure}[h]
	\includegraphics[height=65mm,width=1\columnwidth]{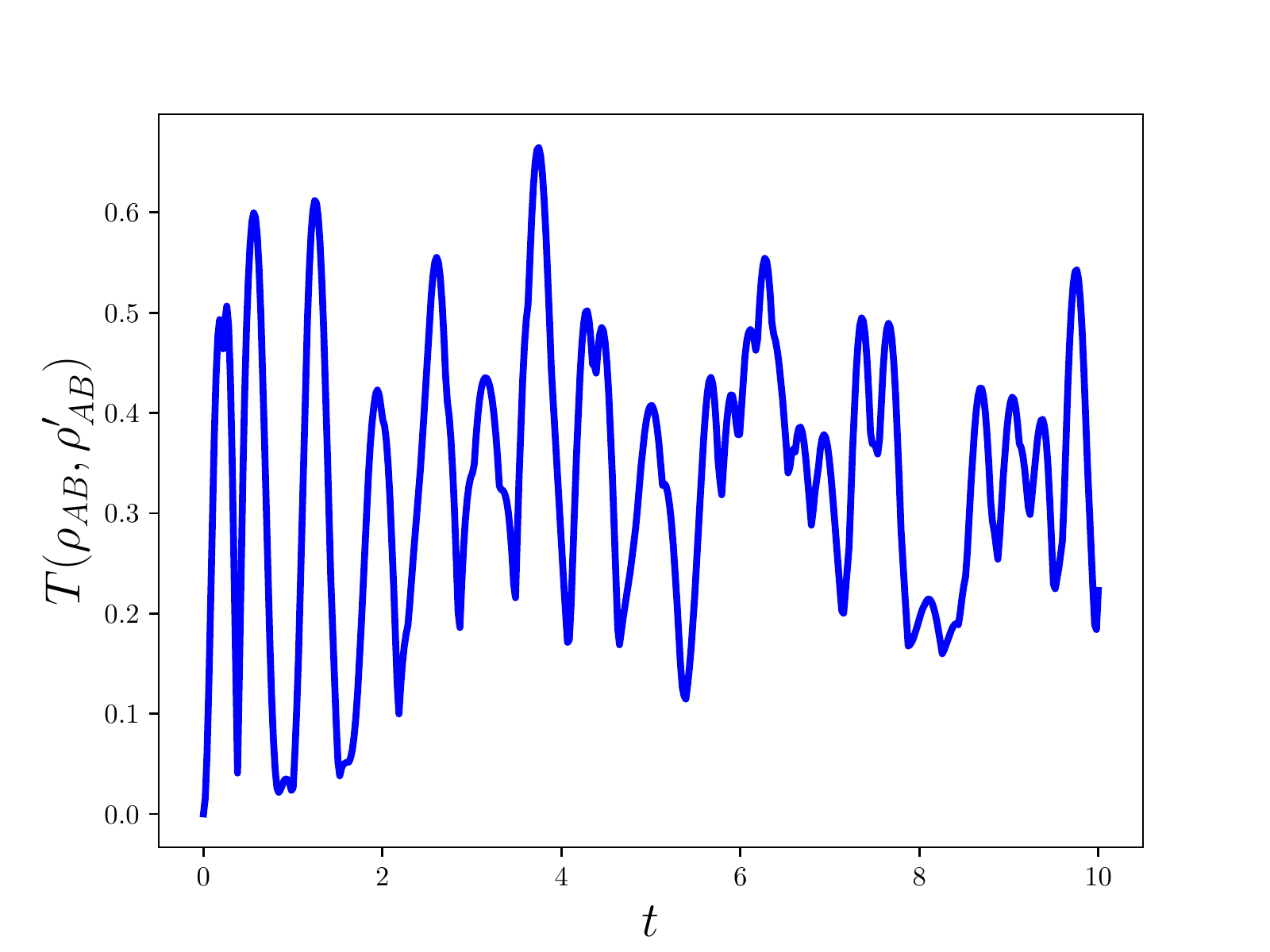}
	\caption{Variation of trace distance ($T(\rho_{AB}, \rho'_{AB})$) between two-qubit central spin systems, in local bath scenario, when interaction in bath spins is turned on $(\rho_{AB})$ or off $(\rho'_{AB})$. The parameters have following values: $\epsilon_1 = 2.4$, $\epsilon_2 = 2.5$, $\omega_1 = 3.0$, $\omega_2 = 3.1$, $\omega_a = 2.0$, $\omega_b = 2.1$, $M=25$  $N=25$, $T=1$.}
	\label{fig-tr-dist-2-qubit-local}
\end{figure}
Trace distance plotted in Fig. \ref{fig-tr-dist-2-qubit-local} again shows the difference in the system's evolution caused by the interaction among the spins of the individual baths. A comparison between the above figures reveals that while the trace distance can be zero at certain times for the single central spin model and also for the two-qubit central spin model in the global bath scenario, it is not so for the two-qubit central spin model in the local bath regime.
\subsection{von Neumann entropy}
For a given state $\rho$ of a quantum system, the von Neumann entropy of the system is given as 
\begin{equation}
    S(\rho) = -\mathrm{Tr}(\rho\log\rho).
\end{equation}
In the single spin case, taking the initial state to be $\ket{1}$, the von Neumann entropy of the states is calculated using  Eq. (\ref{single-qubit-rho-case1}) and Eq. (\ref{single-qubit-Kraus-case2}). 
\begin{figure}[h]
	\includegraphics[height=65mm,width=1\columnwidth]{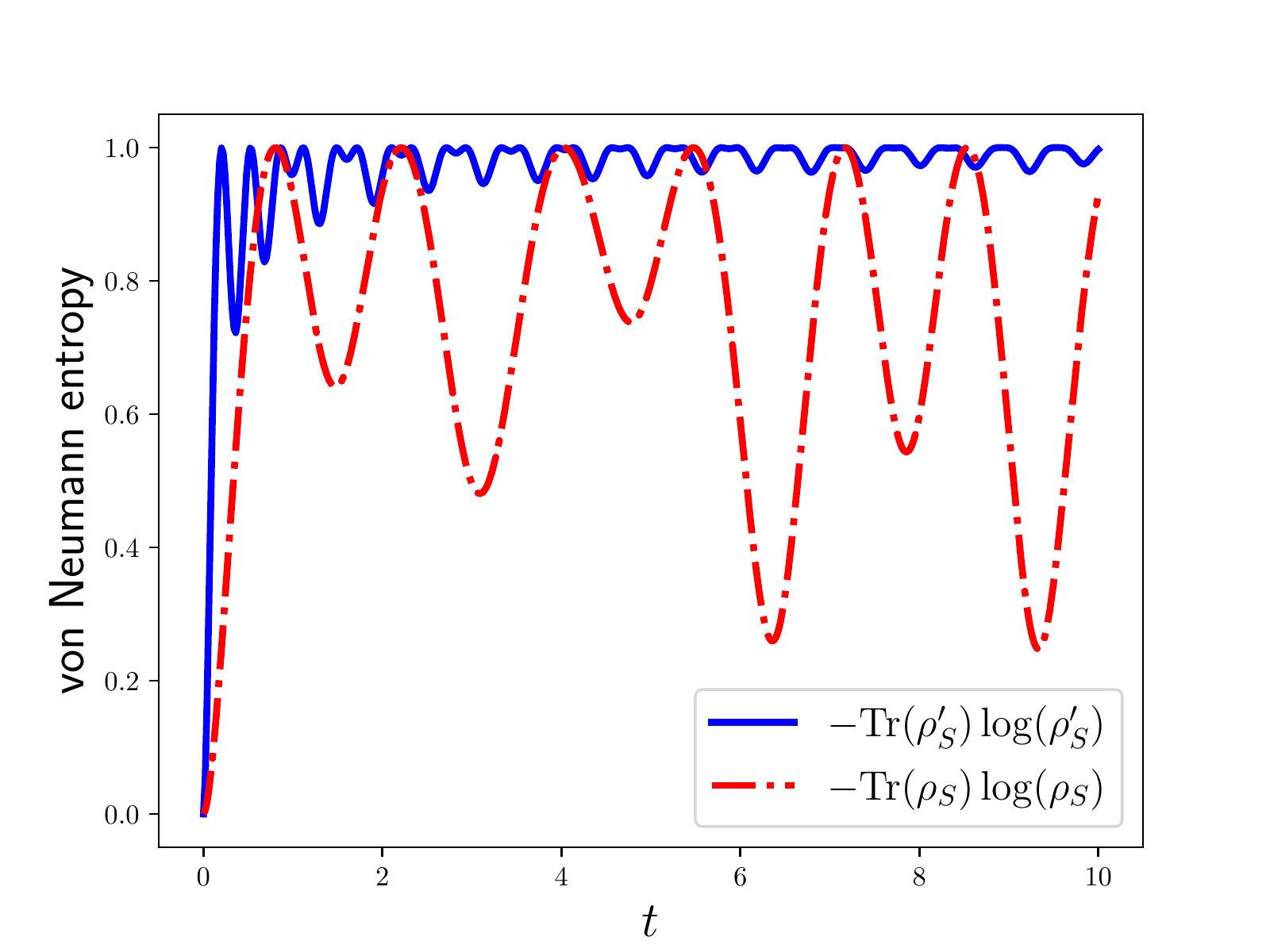}
	\caption{Variation of von Neumann entropy with time in single qubit central spin model. The solid blue line corresponds to the case where bath spins do not interact among themselves, and the dot-dashed red line corresponds to the case where bath spins interact with each other. The parameters have following values: $\epsilon = 1$, $\omega = 2$, $\omega_0 = 2$, $N=100$, $T=1$.}
	\label{fig-von-ent_single}
\end{figure}
Figure \ref{fig-von-ent_single} depicts the difference in the evolution of von Neumann entropy for the single qubit central spin model for (a) where the bath spins interact and (b) where they do not interact with each other. For the non-interacting bath spins scenario, the von Neumann entropy quickly reaches the maximum value, with small fluctuations; the evolution takes the state close to a maximally mixed state. In contrast, for the interacting bath spins, we see a larger variation in the von Neumann entropy depicting a larger exchange of information between the bath and the central spin. 

In the two-qubit case, we consider the initial state of the central spin system to be $\ket{11}$ and calculate the von Neumann entropy for both local bath and global bath scenarios. 
\begin{figure}[h]
	\includegraphics[height=80mm,width=1\columnwidth]{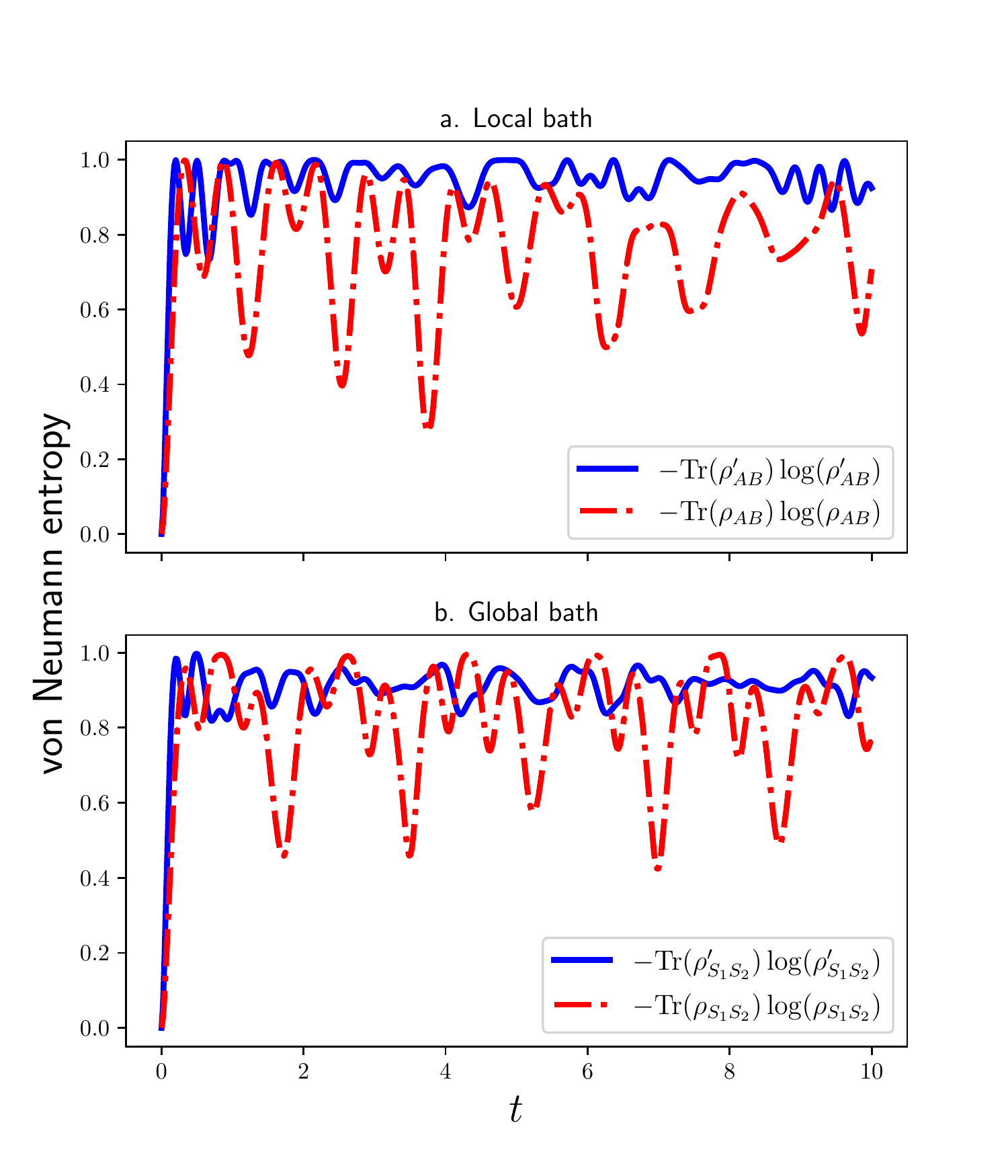}
	\caption{Variation of von Neumann entropy with time in the two-qubit central spin model. Subplot (a) refers to the local bath scenario and (b) refers to the global bath scenario. The solid blue line corresponds to the case where bath spins do not interact among themselves, and the dot-dashed red line corresponds to the case where bath spins interact with each other. The parameters are: $\epsilon_1 = 2.4$, $\epsilon_2 = 2.5$, $\omega_1 = 3.0$, $\omega_2 = 3.1$, $\omega_a = 2.0$, $\omega_b = 2.1$, $\delta = 4$, $M=15$  $N=15$, $T=1$. }
	\label{fig-von-2-qubit-local-and-global}
\end{figure}
Figure \ref{fig-von-2-qubit-local-and-global}(a) depicts the von Neumann entropy for the local bath scenario while 
the Fig. \ref{fig-von-2-qubit-local-and-global}(b) illustrates the global bath scenario.
For both cases, we see a greater variation of von Neumann entropy in the case of interacting bath spins. The variation in the von Neumann entropy approaches a fixed value when bath spins are non-interacting. 

Considering the bipartition between the central and the bath spins, it is observed that the entanglement between them, in the case of non-interacting bath spins, is higher. This is because partial tracing of the bath degrees of freedom in the non-interacting bath spins case results in the reduced system approaching a maximally mixed state. We argue that in the case of interacting bath spins, the correlation is shared between bath spins as well as with the central spin. Therefore, the entanglement between the central and bath spins is less here, as depicted by the von-Neumann entropy. In contrast, in the non-interacting bath spins scenario, the central spin directly shares correlations with the bath spins.
Comparing the blue and red curves separately in Figs. \ref{fig-von-2-qubit-local-and-global}(a) and \ref{fig-von-2-qubit-local-and-global}(b), we observe that in the case of global bath, the frequency of the red curve is greater because of the sharing of correlations between the bath and central spins. In the case of the blue curve, the behaviour is similar for both local and global baths.

\section{\label{sec:QSL-time}Quantum speed limit time}
Mandelstam and Tamm (MT)~ and Margolus and Levitin (ML)-type bounds on speed limit time~\cite{mandelstam1945, MARGOLUS} are estimated by using the geometric approach, using the Bures angle, to quantify the closeness between the  initial and final states. 
This approach was used to provide a bound, for the initial pure state $\rho_0=\vert\psi_0\rangle\langle\psi_0\vert$, on the quantum speed limit time $(\tau_{QSL})$~\cite{deffner2013quantum} as 
\begin{equation}
    \tau_{QSL}=\max\Bigg\{\frac{1}{\Lambda^{\textrm{op}}_{\tau}},\frac{1}{\Lambda^{\textrm{tr}}_{\tau}},\frac{1}{\Lambda^{\textrm{hs}}_{\tau}}\Bigg\} \sin^2[\mathcal{B}],
    \label{tau-qsl}
\end{equation}
where  $\mathcal{B}(\rho_0,\rho_t)=\arccos(\sqrt{\langle\psi_0\vert\rho_t\vert\psi_0\rangle})$, and ${\Lambda^{\textrm{op}}_{\tau}}$,${\Lambda^{\textrm{tr}}_{\tau}}$, and ${\Lambda^{\textrm{hs}}_{\tau}}$ are the operator, Hilbert-Schmidt and trace norms, respectively,
\begin{equation}
    \Lambda^{\textrm{op,tr,hs}}_{\tau}=\frac{1}{\tau}\int^{\tau}_{0} dt \vert\vert \mathcal{L}(\rho_t)\vert\vert_{\textrm{op,tr,hs}}.
    \label{B_spdlmt_1}
\end{equation}  
The three norms ${\Lambda^{\textrm{op}}_{\tau}}$, ${\Lambda^{\textrm{tr}}_{\tau}}$, and ${\Lambda^{\textrm{hs}}_{\tau}}$ are the operator, trace and Hilbert-Schmidt norms, respectively. From the norm inequalities, it can be shown that the operator norm of the generator provides a tighter bound on the quantum speed limit time.
Using Eq. (\ref{tau-qsl}), we estimate the speed limit time for the dynamics of interacting and non-interacting central spin systems, and their behaviour is studied for different sizes of bath spins. We also discuss the connection between quantum speed limit time and quantum correlations for local and global bath effects.  To begin with, we consider the dynamics of $\tau_{QSL}$ for the single qubit central spin model for an initial state  $\ket{1}$. 
\begin{figure}[h]
	\includegraphics[height=80mm,width=1\columnwidth]{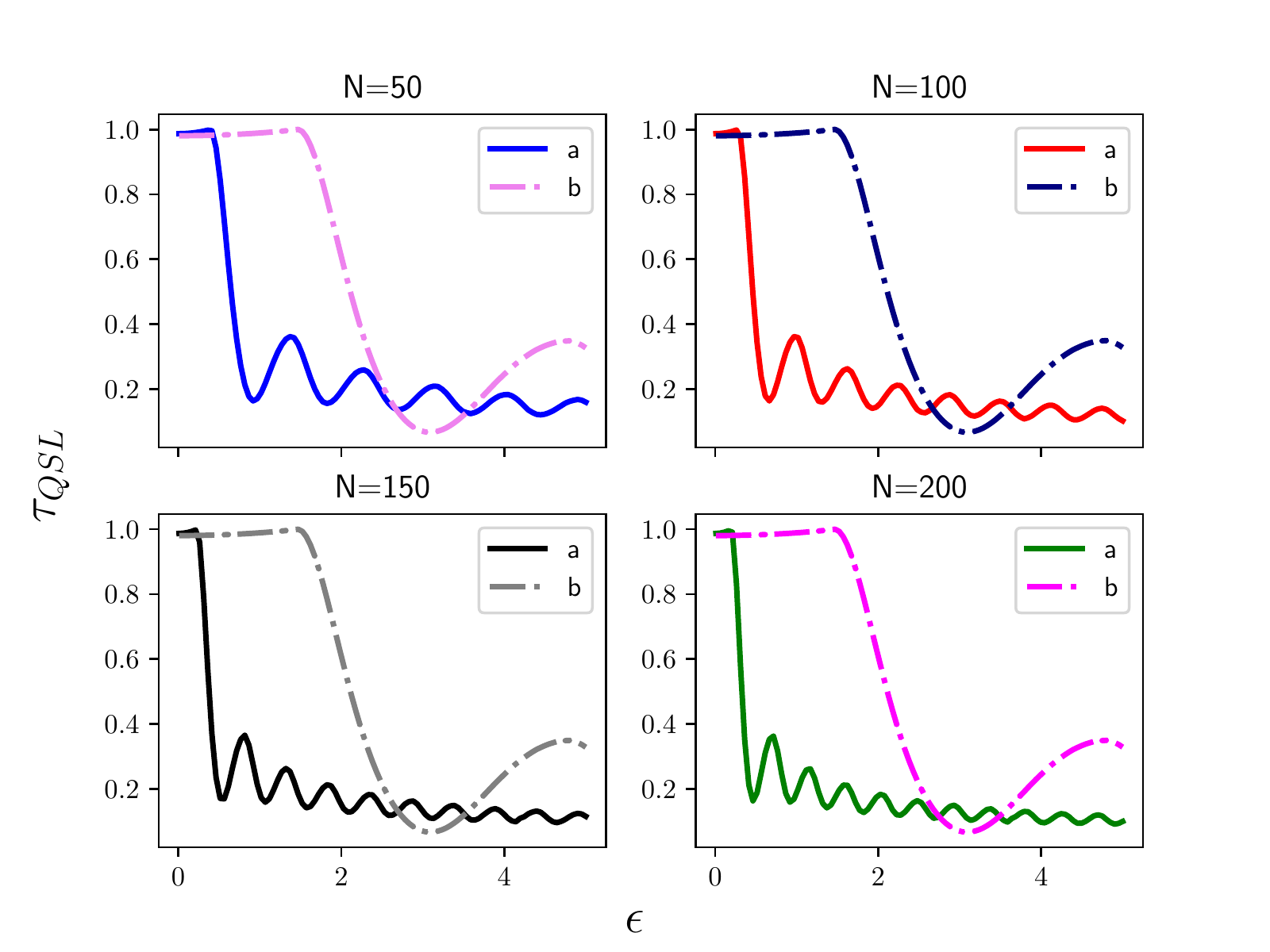}
	\caption{Quantum  speed limit time $\tau_{QSL}$ with the central spin and bath interaction parameter, $\epsilon$ for different numbers of bath spins $N$ for a single qubit central spin system. In each subplot, (a) corresponds to the case when bath spins do not interact with each other (Eq. (\ref{single-qubit-Kraus-case2})) and (b) corresponds to the case when bath spins interact with each other (Eq. (\ref{single-qubit-rho-case1})). The parameters are chosen to be: $\omega = 2$, $\omega_0 = 2$, $\tau=1$ and $T=1$.}
	\label{qsl-1-qubit-vary-e}
\end{figure}%
The effect of the central spin-bath interaction parameter ($\epsilon$) on QSL time for a single qubit central spin system can be seen in Fig. \ref{qsl-1-qubit-vary-e}. In the case of non-interacting bath spins, the frequency of oscillation of $\tau_{QSL}$ rapidly increases on increasing the number of bath spins. However, when the bath spins are interacting, the behaviour of QSL time is nearly the same for different numbers of spins in the bath.
\begin{figure}[h]
	\includegraphics[height=80mm,width=1\columnwidth]{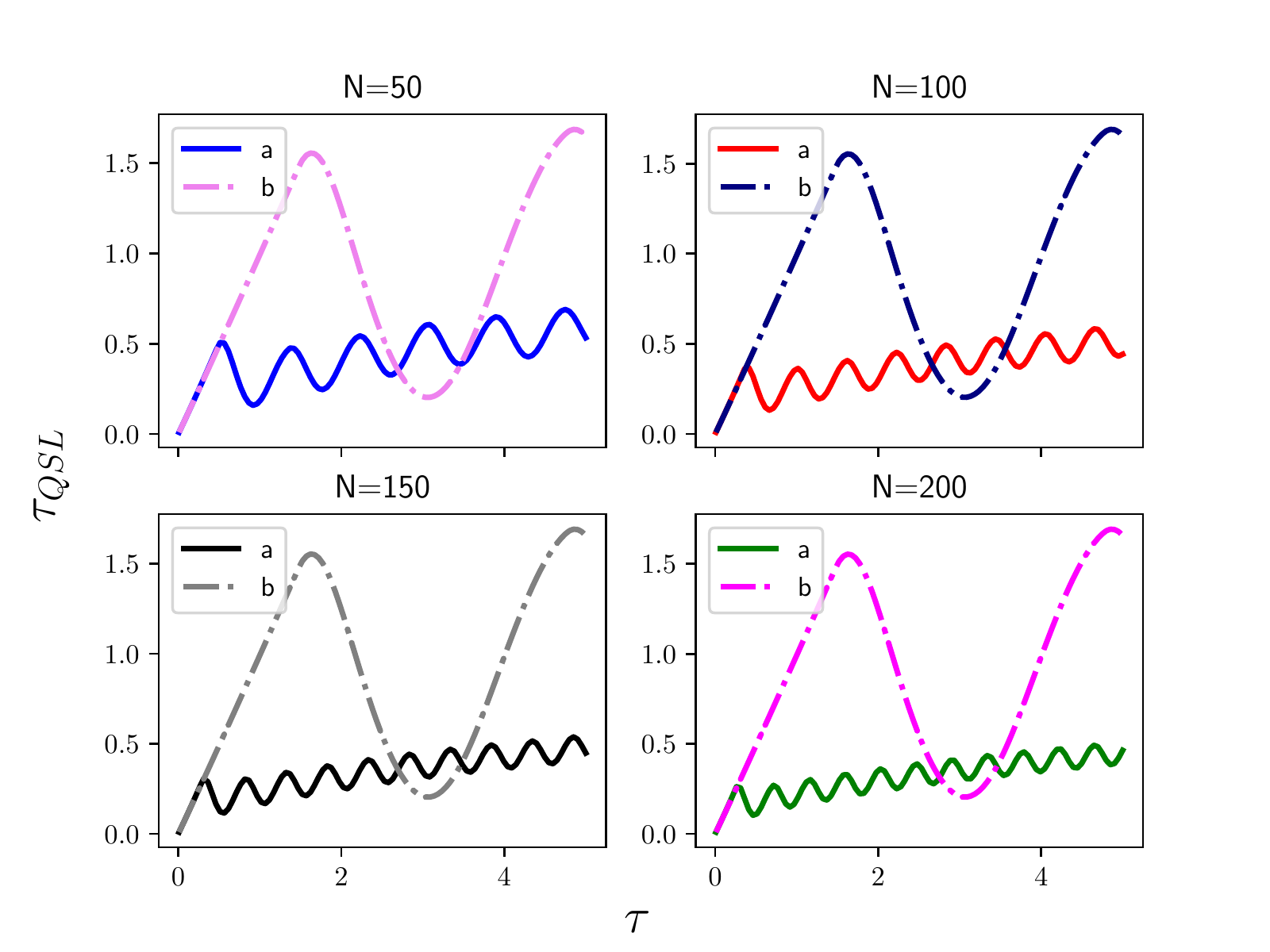}
	\caption{Variation of the quantum speed limit time $\tau_{QSL}$ with the driving time $\tau$ for different numbers of bath spins $(N)$ for a single qubit central spin system. In each subplot, (a) corresponds to the case when bath spins do not interact with each other (Eq. (\ref{single-qubit-Kraus-case2})) and (b) corresponds to the case when bath spins interact with each other (Eq. (\ref{single-qubit-rho-case1})). The parameters have following values: $\epsilon=1$, $\omega = 2$, $\omega_0 = 2$ and $T=1$.}
	\label{qsl-1-qubit-vary-t}
\end{figure}%
From Fig. \ref{qsl-1-qubit-vary-t}, we see that as we increase the size of the bath, in the case of non-interacting bath spins, the QSL time decreases; in other words, the evolution of the system between two states is faster. In the case of interacting bath spins, we observe a minimal change in the QSL time as we increase the number of bath spins. We assert that in this case, due to an additional burden of shared quantum correlations between the bath spins, the system becomes {\it sluggish}, and the results are seen to be nearly the same for different $N$.
The $\tau_{QSL}$ for two-qubit central spin systems is studied below in conjunction with the quantum correlations. 

\section{\label{sec:quantum-correlations}Quantum correlations and connection to QSL time}
Entanglement is one of the most important resources in quantum information. Concurrence is an entanglement measure for a two-qubit system \cite{concurrence1, Wootters2001EntanglementOF}, defined as
\begin{equation}
	{\mathcal C}(\rho_{PQ}) = \max{\{0, \lambda_1-\lambda_2-\lambda_3-\lambda_4\}}, 
	\label{concurrence}
\end{equation}
where $\lambda_i$'s are the eigenvalues of the matrix $\sqrt{\sqrt{\rho_{PQ}}{\tilde \rho}_{PQ}\sqrt{\rho_{PQ}}}$, in decreasing order, and ${\tilde\rho}_{PQ} = (\sigma^y\otimes\sigma^y)\rho_{PQ}^*(\sigma^y\otimes\sigma^y)$. We consider the initial state of the two-qubit central spin systems to be $\frac{1}{\sqrt{2}}(\ket{00}+\ket{11})$ and calculate the concurrence for the evolution of the system in the local and global bath scenarios and compare it with the quantum speed limit time $(\tau_{QSL})$, as given below. 
Another well-known measure of  quantum correlation in a quantum system is quantum discord \citep{quantum-discord-Zurek, quantum-discord-dev, discord2, discord3}. Here quantum discord between two central spins ($P$ and $Q$) is given by
\begin{equation}
	\mathcal{D}(\rho_{PQ}) = \mathcal{S}(\rho_P) - \mathcal{S}(\rho_{PQ}) + \mathcal{S}(\rho_{P|Q}),
	\label{discord-definition}
\end{equation}
where $\mathcal{S}({\rho_{Q}})$ is the von Neumann entropy of the subsystem of central spin $Q$, $\mathcal{S}(\rho_{PQ})$ is the joint von-Neumann entropy of the system $\rho_{PQ}$ and $\mathcal{S}(\rho_{P|Q})$ is the quantum conditional entropy of the system. Further, the quantum conditional entropy is given by 
\begin{equation}
	\mathcal{S}(\rho_{P|Q}) = \min_{\{M_j\}} \sum_{j=1}^2 p_j\mathcal{S}(\rho_{P|M_j}),
	\label{quant-cond-ent}
\end{equation}
where $\rho_{P|M_j}$ is the reduced state of the subsystem $P$ after a measurement is performed over subsystem $Q$ and $M_j$'s are the generalized measurement operators for subsystem $Q$. For single qubit, these measurement operators are given by 
\begin{equation}
	\begin{aligned}
		M_1 =\ketbra{u_Q}{u_Q}, && M_2 =\ketbra{v_Q}{v_Q},
		\label{discord-measurement-operators}
	\end{aligned}
\end{equation}
where $\ket{u_Q} = \cos(\theta)\ket{0} + e^{i\phi}\sin(\theta)\ket{1}$, and $\ket{v_Q} = \sin(\theta)\ket{0} - e^{i\phi}\cos(\theta)\ket{1}$, such that $0\le\theta\le\pi/2$ and $0\le\phi\le 2\pi$. 
We calculate the quantum discord for the evolution of the two-qubit central spin systems, from the initial state $\frac{1}{\sqrt{2}}(\ket{00}+\ket{11})$, for both the local and global bath scenarios and compare it with the concurrence and quantum speed limit time $(\tau_{QSL})$, as given below.

\subsection{\label{sec:conect-qcorr-qsl}Connection between quantum correlations and QSL time}
We first consider the dynamics of the two-qubit central spin system in the local bath scenario. We use Eq. (\ref{tau-qsl}) and dynamics of the state defined in Eq. (\ref{reduced-state-local-bath}) to calculate the QSL time of the system, in the case of the individual baths of interacting spins and Eq. (\ref{reduced-state-local-case2}) to calculate the QSL time in case of individual baths of non-interacting spins. Further, we analyze the dynamics of the quantum correlations in both cases, in particular, entanglement and quantum discord.
\begin{figure}[h]
	\includegraphics[height=80mm,width=1\columnwidth]{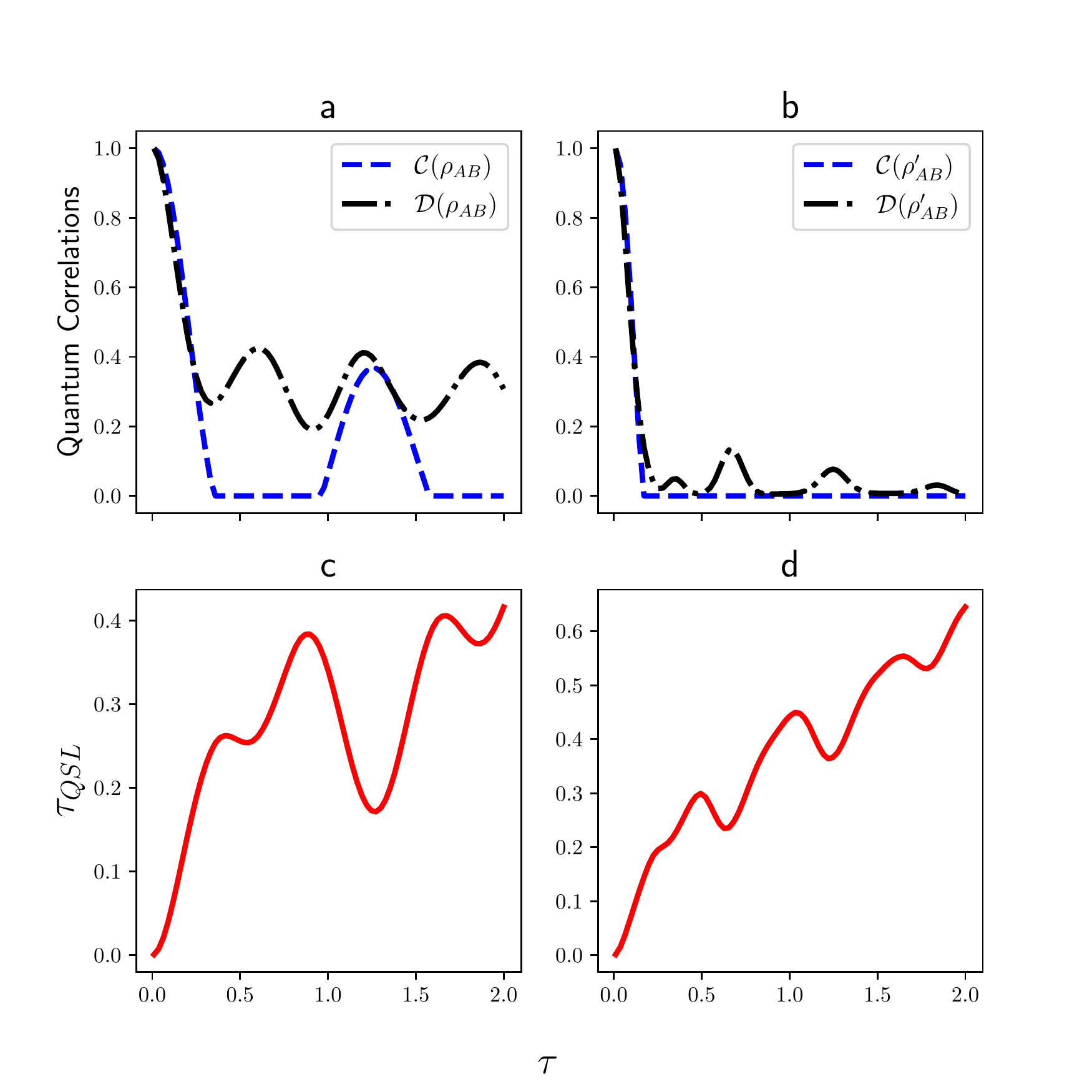}
	\caption{Variation of the QSL time ($\tau_{QSL}$) and quantum correlations (Concurrence ${\mathcal C}$ and discord ${\mathcal D}$), in local bath scenario, for the two-qubit central spin model with time. In (a) and (c), the spins of the individual baths interact with each other (Eq. (\ref{reduced-state-local-bath})) while in (b) and (d), they do not interact with each other (Eq. (\ref{reduced-state-local-case2})). The parameters are: $\epsilon_1 = 2.5$, $\epsilon_2 = 2.5$, $\omega_1 = 3.0$, $\omega_2 = 3.0$, $\omega_a = 2.0$, $\omega_b = 2.0$, $M=15$  $N=15$, $T=1$.}
	\label{local_2_qubit_qcorr_qsl_with_tau}
\end{figure}%
Figure \ref{local_2_qubit_qcorr_qsl_with_tau} depicts the changes in QSL time and quantum correlations \textit{w.r.t.} time for the two-qubit central spin model in local bath scenario. We observe that the dynamics of the quantum correlations are in contrast with the changes in QSL time for both the cases of interacting and non-interacting individual bath spins. This suggests that the speed of the system's evolution is aided by increments in the quantum correlations in that the blips in quantum correlations correspond to the dips in QSL time. This is consistent with the behaviour between QSL time and quantum correlations for non-unital channels, which is the case here for the model studied, and as pointed out in Ref.~\cite{paulson2021hierarchy}.

We next consider the case of the two-qubit central spin model in the global bath scenario defined by the Hamiltonians, Eqs. (\ref{global-hamiltonian-case1}) and (\ref{global-hamiltonian-case2}), respectively. We analyze the dynamics of QSL time and quantum correlations for a given central spin-spin interaction strength ($\delta$) with time. 
\begin{figure}[h]
	\includegraphics[height=80mm,width=1\columnwidth]{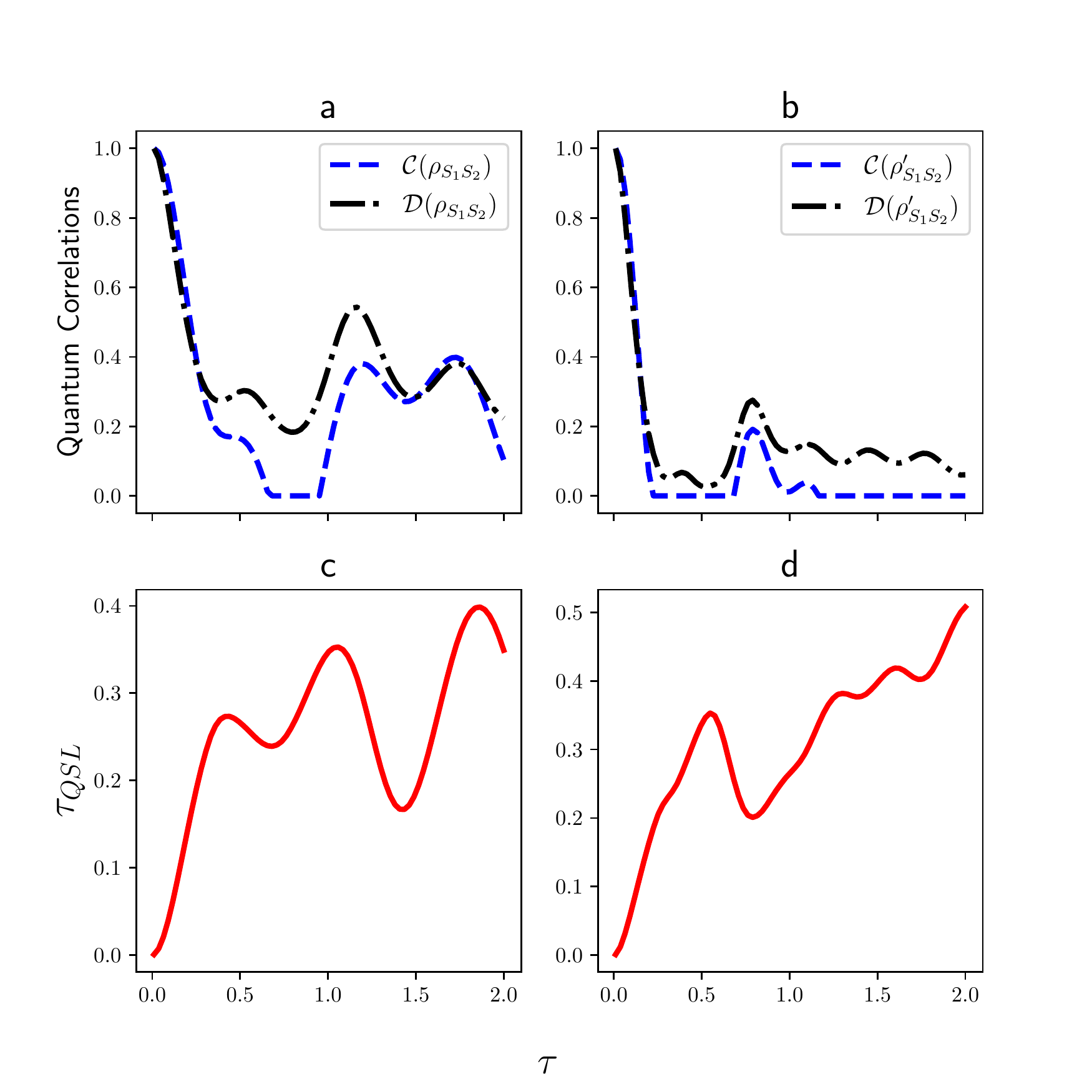}
	\caption{ QSL time ($\tau_{QSL}$) and quantum correlations (Concurrence ${\mathcal C}$ and Discord ${\mathcal D}$), in global bath scenario, for the two-qubit central spin model with time. In (a) and (c), the spins of the individual baths interact with each other (Eq. (\ref{rho-global-case1})) while in (b) and (d), they do not interact with each other (Eq. (\ref{rho-global-case2})). The parameters are chosen to be: $\omega_1 = 3.0$, $\omega_2 = 3.1$, $\omega_a = 2.0$, $\omega_b = 2.1$, $\delta = 4$, $\epsilon_1 = 2.4$, $\epsilon_2 = 2.5$, $M=10$, $N=10$ and $T=1$.}
	\label{global_2_qubit_qcorr_qsl_with_tau}
\end{figure}
In Fig. \ref{global_2_qubit_qcorr_qsl_with_tau}, we witness that the changes in the dynamics of quantum correlations are again in contrast with the changes in QSL time for both the cases of interacting and non-interacting individual bath spins. Once again, it is observed that the increment in quantum correlations facilitates the speed of the system's evolution. We next see the impact of the interaction strength ($\delta$) between the two central spins on the QSL time and quantum correlations at a given time.  
\begin{figure}[!h]
	\includegraphics[height=80mm,width=1\columnwidth]{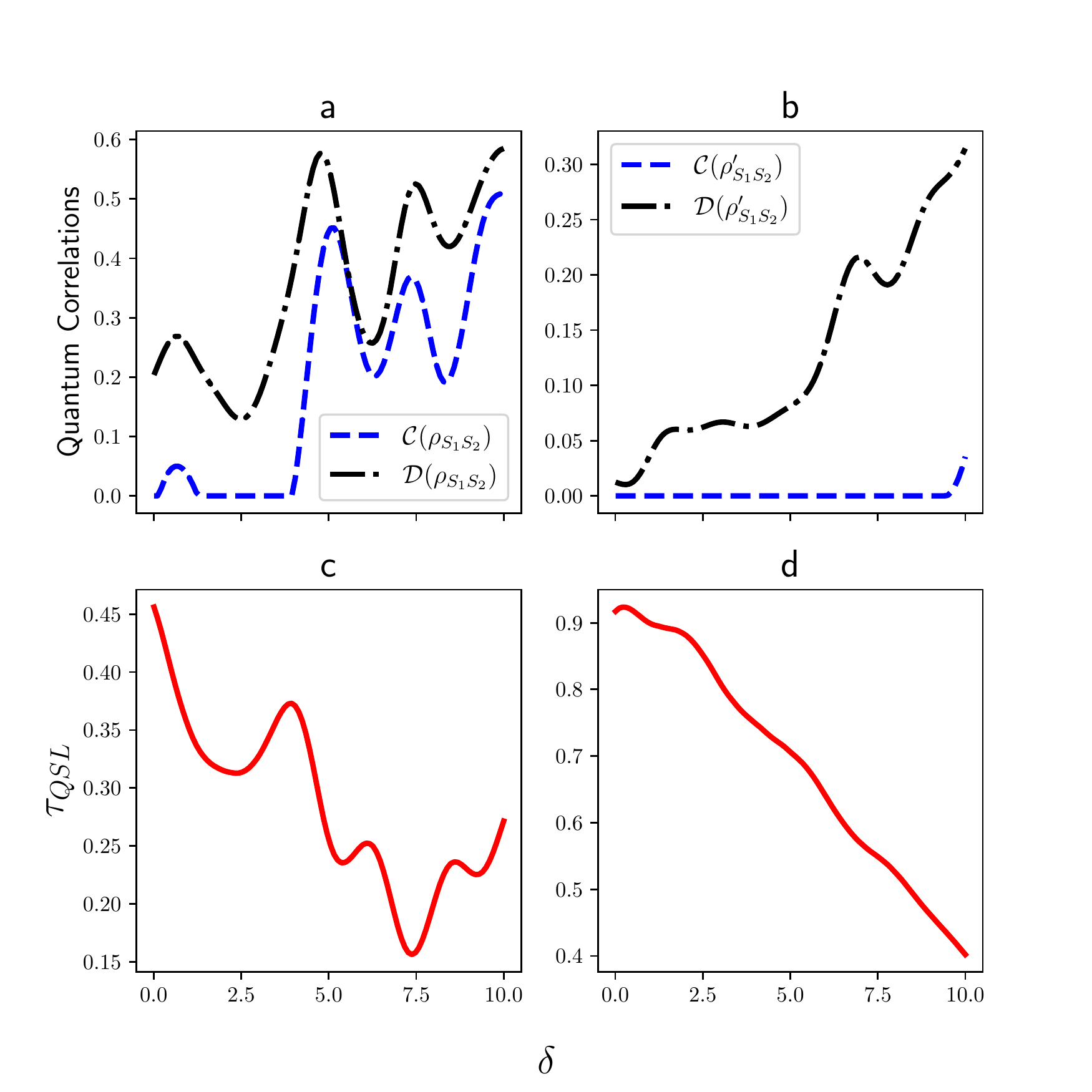}
	\caption{Variation of the QSL time ($\tau_{QSL}$) and quantum correlations (Concurrence ${\mathcal C}$ and discord ${\mathcal D}$), in global bath scenario, for the two-qubit central spin model with the interaction strength between central spins $\delta$. In (a) and (c), the spins of the individual baths interact with each other (Eq. (\ref{rho-global-case1})) while in (b) and (d), they do not interact with each other (Eq. (\ref{rho-global-case2})). The parameters have following values: $\omega_1 = 3.0$, $\omega_2 = 3.1$, $\omega_a = 2.0$, $\omega_b = 2.1$, $\tau = 4$, $\epsilon_1 = 2.4$, $\epsilon_2 = 2.5$, $M=10$, $N=10$ and $T=1$.}
	\label{global_2_qubit_qcorr_qsl}
\end{figure}%
In Fig. \ref{global_2_qubit_qcorr_qsl}, we observe that the quantum correlations in the system are non-monotonically increasing as the interaction strength between the two central spins increases, which is true for both the cases of interacting and non-interacting spins of the individual baths. Here again, the speed of the system's evolution is faster when quantum correlations increase. 
A general remark that can be made from the above figures is that there are regimes in the parameter space where the entanglement is zero, but quantum discord is non-zero. We can see that, in general, interaction in the bath spins aid in the generation of quantum correlations, as can be seen by comparing Figs. \ref{global_2_qubit_qcorr_qsl_with_tau}(a) with \ref{global_2_qubit_qcorr_qsl_with_tau}(b) and \ref{global_2_qubit_qcorr_qsl}(a) with \ref{global_2_qubit_qcorr_qsl}(b). 
The revivals in the quantum correlations in Figs.~\ref{local_2_qubit_qcorr_qsl_with_tau} and ~\ref{global_2_qubit_qcorr_qsl_with_tau}
brings out the non-Markovian nature of the system.
\begin{figure}[h]
	\includegraphics[height=65mm,width=1\columnwidth]{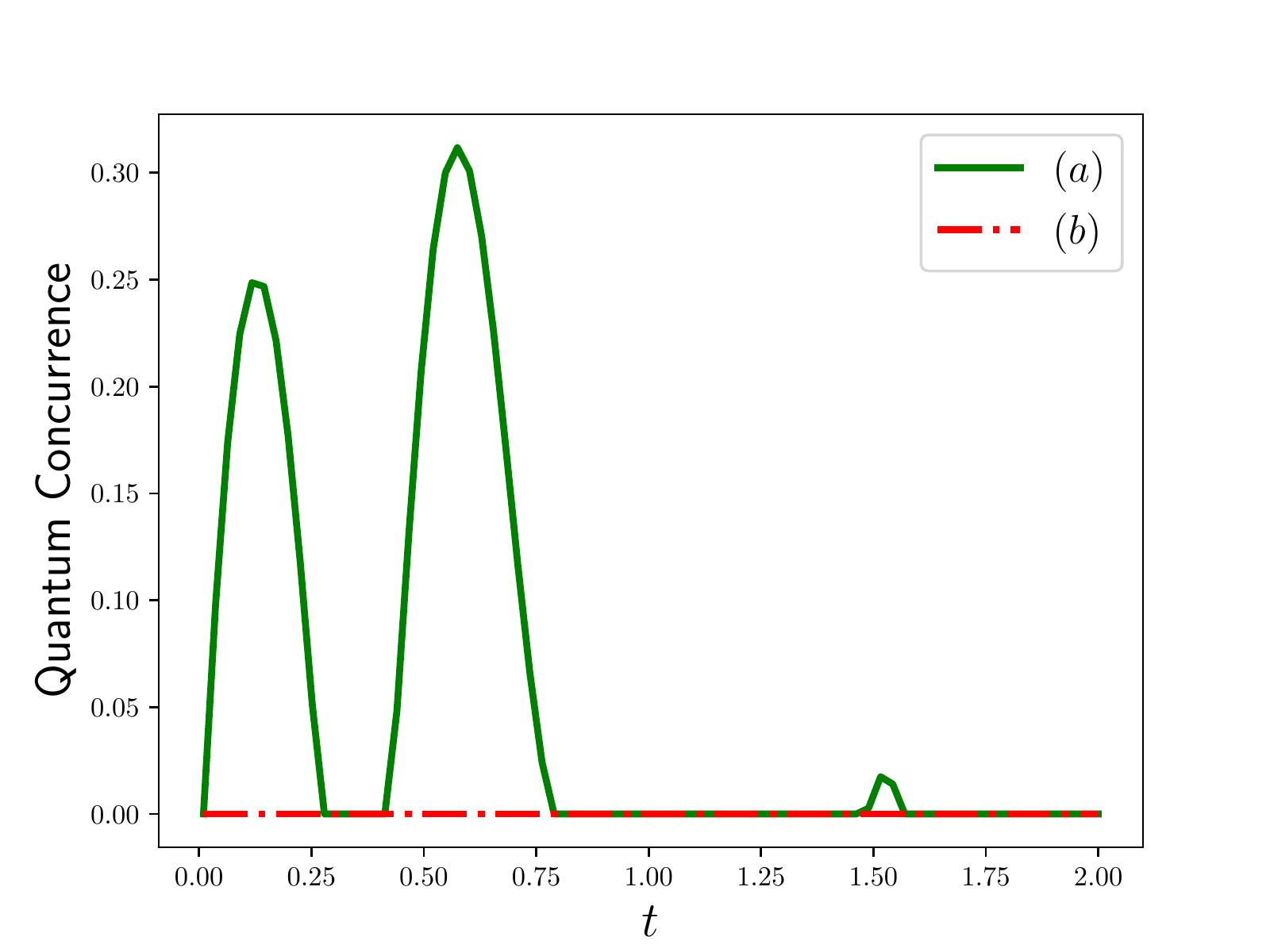}
	\caption{Variation of quantum concurrence ${\mathcal C}$ in the global bath scenario with interacting bath spins for the two-qubit central spin model with time. Here the initial state of the reduced central spin system is chosen to be $\ket{11}$. Here (a) denotes the $\sigma_x-\sigma_x$ interaction and (b) denotes the $\sigma_z-\sigma_z$ interaction between both the central spins. The parameters are chosen to be: $\omega_1 = 3.0$, $\omega_2 = 3.1$, $\omega_a = 2.0$, $\omega_b = 2.1$, $\delta = 4$, $\epsilon_1 = 2.4$, $\epsilon_2 = 2.5$, $M=10$, $N=10$ and $T=0.1$.}
	\label{fig-diff_concurrence_xx_zz_int}
\end{figure}
The evolution of a separable initial state $\ket{11}$ of the two central spins model is depicted in Fig.~\ref{fig-diff_concurrence_xx_zz_int}. For the $\sigma_z-\sigma_z$ interaction, as studied in the present work, the quantum entanglement is seen to be zero at all times. However, if instead we use a $\sigma_x-\sigma_x$ interaction between the two central spins with the same separable initial state, a non-zero entanglement is observed.
\section{\label{conclusion}Conclusions}
Here we have focused on the different types of single-qubit and two-qubit central spin models. Special emphasis is laid on whether the bath spins interact or are non-interacting. For the two-qubit central spin model, we have considered the cases of whether or not the central spins interact with each other. These are called the global or local bath scenarios, respectively. In the literature~\cite{two-qubit-central-spin, mukhopadhyay2017}, the central spin model is solved using the Holstein-Primakoff transformation. Here we have exactly solved the single and two-qubit central spin models numerically and studied the impact of interacting
bath spins.
Trace distance and von Neumann entropy have been used to characterize these models. We have observed that the interaction of bath spins plays a vital role in the system's dynamics. The evolution of the single and two-qubit central spin models, where bath spins are non-interacting, are seen to evolve into a nearly maximally mixed state. The impact of bath size on the QSL time for the single qubit model revealed that in the case of non-interacting bath spins, the QSL time is impacted, but the effect was not prominent for the possibility of interacting bath spins. Further, we have studied the dynamics of quantum correlations, particularly entanglement and quantum discord, in the two-qubit central spin models. A connection was observed between the QSL time and quantum correlations, revealing that the higher quantum correlations can result in a faster system evolution speed.       

\section*{ACKNOWLEDGEMENTS}
SB acknowledges support from the Interdisciplinary Cyber-Physical Systems (ICPS) programme of the Department of Science and Technology (DST), India, Grant No.: DST/ICPS/QuST/Theme-1/2019/6 and DST/ICPS/QuST/Theme-1/2019/13. SB also acknowledges support from the Interdisciplinary Research Platform (IDRP) on Quantum Information and Computation (QIC) at IIT Jodhpur.

	\bibliographystyle{apsrev4-1}
	\bibliography{ref}
	
\end{document}